\documentclass[twoside,english]{iopart}
\usepackage[T1]{fontenc}
\usepackage[latin9]{inputenc}
\usepackage[letterpaper]{geometry}
\usepackage{amstext}
\usepackage{graphicx}
\usepackage{amssymb}

\makeatletter
\usepackage{iopams}
\usepackage{setstack}

\usepackage[numbers, sort&compress]{natbib}

\def\newblock{\hskip .11em plus .33em minus .07em}

\makeatother

\usepackage{babel}

\begin{document}

\title{Substitutional disorder and charge localisation in manganites}

\author{{\Large Eduardo V Castro$^{1,2}$ and J M B Lopes dos Santos}$^{1}$}

\address{{\large $^{1}$ CFP and Departamento de F\'{\i}sica, Faculdade de
Ci\^encias Universidade do Porto, P-4169-007 Porto, Portugal}}

\address{{\large $^{2}$ Instituto de Ciencia de Materiales de Madrid, CSIC,
Cantoblanco, E-28049 Madrid, Spain}}

\ead{{\large {evcastro@fc.up.pt, evcastro@icmm.csic.es}}}
\begin{abstract}
In the manganites $RE{}_{1-x}AE_{x}$MnO$_{3}$ ($RE$ and $AE$ being
rare-earth and alkaline-earth elements, respectively) the random distribution
of $RE{}^{3+}$ and $AE{}^{2+}$ induces random, but correlated shifts
of site energies of charge carriers in the Mn sites. We consider a
realistic model of this diagonal disorder, in addition to the double-exchange
hopping disorder, and investigate the metal-insulator transition as
a function of temperature, across the paramagnetic-ferromagnetic line,
and as a function of doping~$x$. Contrary to previous results, we
find that values of parameters, estimated from the electronic structure
of the manganites, are not incompatible with the possibility of a
disorder induced metal to insulator transition accompanying the ferromagnetic
to paramagnetic transition at intermediate doping ($x\sim0.2-0.4$).
These findings indicate clearly that substitutional disorder has to
be considered as an important effect when addressing the colossal
magnetoresistance properties of manganites.
\end{abstract}


\submitto{\JPCM }

\maketitle

\section{Introduction}

The discovery of colossal magnetoresistance (CMR) in the manganites
$RE{}_{1-x}AE_{x}$MnO$_{3}$ (where $RE$ and $AE$ are trivalent
rare-earth and divalent alkaline-earth ions, respectively) has attracted
much interest to these perovskites \cite{CVvM99,LP00,SJ01,DHM01,Edw02}.
On the one hand, understanding the origin of the CMR effect from a
fundamental point of view is expected to give some insight into the
complex behaviour seen in other strongly correlated systems, as high-temperature
superconductors. On the other, such a colossal response to an external
perturbation still makes these Mn oxides very appealing from the point
of view of applications \cite{Dag05}. This CMR effect, particularly
{}``colossal'' in the so-called intermediate-bandwidth manganites
for doping $x\sim0.2-0.4$ \cite{DHM01}, is intrinsically related
with the presence of metallic behaviour below the Curie temperature
($T_{\textrm{C}}$) and insulating behaviour above it. Understanding
the nature of this metal-insulator transition (MIT) accompanying the
magnetic transition is thus a key point in the comprehension of the
CMR effect. 

Qualitatively, the correlation between transport and magnetic properties
is well understood via Zener's double-exchange (DE) mechanism \cite{ZEN51b,AH55,DeGennes60}:
the spin of itinerant $e_{g}$ and local $t_{2g}$ Mn \emph{d} electrons
are aligned by Hund's rule; to lower the kinetic energy ferromagnetism
is favoured, and at low temperature ($T<T_{\textrm{C}}$) a ferromagnetic
half-metal is realised \cite{CVvM99,Edw02}. Consequently, early proposals
for the MIT focused on the random nature of hopping in the paramagnetic
phase ($T>T_{\textrm{C}}$) \cite{KA88,AKT88,Var96,KAK99}. Quantitative
analysis based on the transfer matrix method showed concomitant ferro-paramagnetic
and MITs for $x\sim0.2-0.4$ when diagonal disorder is added to the
DE model \cite{LZB+97,SXS+97b}. The resultant CMR effect \cite{SXS+97a,ZMJ05},
however, requires an amount of diagonal disorder that seems incompatible
with potential fluctuations originated by the random distribution
of $RE^{3+}$ and $AE^{2+}$ ions \cite{PS97,Edw02}. This apparent
failure of Anderson localisation theories favoured models based on
polaronic formation \cite{MLS95,MSM96,RZB96}, owing to the electron-phonon
coupling due to Jahn-Teller effect in manganites \cite{Edw02}. It
has been argued, however, that manganites fall into an intermediate
electron-phonon coupling regime where small-polarons -- key ingredients
for a MIT driven by electron-phonon interaction -- are hardly formed
in the paramagnetic phase \cite{Edw02,BSD+06}. More recent theories
explain the CMR as an effect of competing orders: in brief, a ferromagnetic
metal competes with a charge-order insulator phase, producing, for
$T\gtrsim T_{\textrm{C}}$, an inhomogeneous state highly sensitive
to external perturbations where CMR is observed \cite{MMF+00,BMM+01,BMD04,MFN03,MFN05,KM06,SAD07}.
The underlying model producing such a phase competition scenario takes
into account, on the same footing, the DE mechanism and the electron-lattice
coupling and, surprisingly, intrinsic disorder in manganites. Adding
quenched disorder has been shown to make the inhomogeneous state even
more sensitive to external perturbations, enhancing the CMR effect,
and avoiding fine tuning of model parameters \cite{SAA+06}. However,
the need to include disorder on the same level as the DE mechanism
and coupling to Jahn-Teller phonons strongly contradicts the observation
that disorder in manganites is sufficiently weak for a virtual crystal
approximation to be reasonable \cite{PS97,Edw02}.

In this paper we consider a realistic model for diagonal (substitutional)
disorder in manganites, in addition to the DE hopping disorder, and
investigate the MIT as a function of temperature, across the paramagnetic-ferromagnetic
line, and as a function of doping~$x$. Contrary to previous results
\cite{Edw02}, we find that values of parameters, estimated from the
electronic structure of manganites \cite{PS96,PS97}, are not incompatible
with the possibility of a disorder induced MIT accompanying the ferromagnetic
to paramagnetic transition at intermediate doping ($x\sim0.2-0.4$).
Therefore, substitutional disorder has to be considered at least on
the same foot as the coupling to the lattice when addressing the CMR
properties of manganites. These findings give support to theories
where disorder is a key ingredient \cite{NV01}, as the phase-competition
scenario for CMR, where, as mentioned above, diagonal disorder plays
a crucial role \cite{MMF+00,BMM+01,BMD04,MFN03,MFN05,SAA+06,KM06,SAD07}.

\section{Model}

In order to model substitutional disorder in manganites we note that,
for each carrier introduced in the system, there is a $RE^{3+}\to AE^{2+}$
substitution. The corresponding change in the Coulomb field shifts
the site energy of an electron in a manganese site at a distance $R$
by\begin{equation}
V(R)=\frac{e^{2}}{4\pi\varepsilon_{0}\varepsilon R},\label{eq:Vmang}\end{equation}
where $\varepsilon$ is the relative dielectric constant of the material.
We can take this effect into account by including a random site energy
term in the DE Hamiltonian,\begin{equation}
H=-\sum_{\left\langle ij\right\rangle }\Bigl(t\left(\mathbf{S}_{i},\mathbf{S}_{j}\right)c_{i}^{\dagger}c_{j}+\mbox{h.c.}\Bigr)+\sum_{i}\epsilon_{i}c_{i}^{\dagger}c_{i}.\label{eq:HDEdis}\end{equation}
The first term on the right hand side in~(\ref{eq:HDEdis}) is the
usual infinite Hund coupling DE Hamiltonian, where the hopping of
$e_{g}$ electrons between nearest-neighbours Mn sites depends on the
background configuration of classical $t_{2g}$ core spins $\mathbf{S}_{i}=S(\sin\theta_{i}\cos\phi_{i},\sin\theta_{i}\sin\phi_{i},\cos\theta_{i})$,
with \begin{equation}
t(\mathbf{S}_{i},\mathbf{S}_{j})=t[\cos(\theta_{i}/2)\cos(\theta_{j}/2)+e^{-i\left(\phi_{i}-\phi_{j}\right)}\sin(\theta_{i}/2)\sin(\theta_{j}/2)].\label{eq:tSiSj}\end{equation}

The second term on the right hand side of~(\ref{eq:HDEdis}) stands
for the diagonal site disorder. It has been modelled in previous works
with a uniform probability distribution for $-W/2\le\epsilon_{i}\le W/2$
(Anderson disorder) \cite{LZB+97,SXS+97a,SXS+97b}. Through the analysis
of the mobility edge trajectory in the energy vs disorder ($W$) plane,
obtained using the transfer matrix method \cite{MK81,MK83,KM93},
it has been found that a MIT occurs when the system crosses the ferro-paramagnetic
transition line, for $0.2<x<0.5$ provided the diagonal disorder is
strong enough, $12t<W<16.5t$. The plausibility of such large value
of the disorder parameter, however, has been questioned \cite{Edw02}
mainly on the basis of density functional results obtained by Pickett
and Singh \cite{PS96,PS97}. In the following we show that a thorough
analysis of the results of \cite{PS97} and \cite{PS96} are, in fact,
not incompatible with disorder in the range $12t<W<16.5t$.

\subsection{Substitutional disorder strength}

Pickett and Singh \cite{PS96,PS97} looked at the $x=1/3$ concentration,
and performed LDA calculations of band structure for a periodic structure
of $\mathrm{La}_{2}\mathrm{CaMn}_{3}\mathrm{O}_{9}$ with a tetragonal
unit cell containing a La--Ca--La set of planes. There are two inequivalent
Mn sites in this structure, one with eight $\mathrm{La}^{3+}$ and
the other with four $\mathrm{Ca}^{2+}$ and four $\mathrm{La}^{3+}$
nearest-neighbours (NNs). The local density of states at the Mn sites
showed a difference of $\Delta\epsilon_{\mathrm{Mn}}\approx0.5\,\mathrm{eV}$
between the band edges for these two types of sites, which was interpreted
as arising from the different charges $\mathrm{Ca}^{2+}$ and $\mathrm{La}^{3+}$.
Thus, $\Delta\epsilon_{\mathrm{Mn}}=4V_{1}$ with $V_{1}$ given by
(\ref{eq:Vmang}) with $R=a\sqrt{3}/2$ (the first NN $\mathrm{La-Mn}$
distance), \begin{equation}
V_{1}=\frac{2e^{2}}{\sqrt{3}\pi\varepsilon_{0}\varepsilon a},\label{eq:V1mang}\end{equation}
where $a\approx3.9\,\mbox{\AA}$ is the $\textrm{Mn}-\mathrm{Mn}$
distance \cite{CVR+95}. From the calculated value of $\Delta\epsilon_{\mathrm{Mn}}$
a dielectric constant $\varepsilon\approx34$ is obtained, or equivalently
$V_{1}\approx0.13\,\mathrm{eV}\approx0.6t$ ($t\approx0.2\,\mbox{eV}$
was used \cite{DHM01}).

Such a dielectric constant, however, is quite unlikely. We should
note that equation~(\ref{eq:V1mang}) is actually a microscopic description,
where $R=a\sqrt{3}/2\approx3.4\,\textrm{\AA}$. Neglecting metallic
screening, we should get a relative permittivity reflecting the polarizability
of the $\mbox{Mn }d-\mbox{O }p$ complex, as pointed out in \cite{PS97}.
Infrared reflectivity measurements on La$_{0.67}$Ca$_{0.33}$MnO$_{3}$
give a high-frequency dielectric constant $\varepsilon_{\infty}\approx7.5$
at $78\,\mbox{K}$ \cite{BKB+99}, which, though being only a lower
bound, casts serious doubts on $\varepsilon\approx34$. On the other
hand, note that the result $\Delta\epsilon_{\mathrm{Mn}}=4V_{1}$
is a special case where only first NNs (La/Ca sites) contribute to
the local potential. A more realistic situation should account for
next NNs contributions.

\begin{figure}
\begin{centering}
\includegraphics[clip,width=0.55\columnwidth]{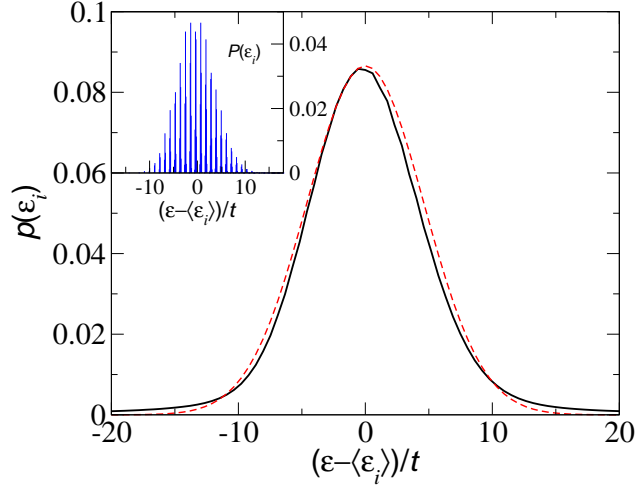}
\par\end{centering}

\caption{\label{fig:dist2NN}Full line shows the probability distribution of
Mn-site energies due to random placement of $RE{}^{3+}$ ($2/3$ probability)
and $AE{}^{2+}$ ($1/3$ probability) on first and second neighbour
sites, obtained by substituting $\delta-$functions in (\ref{eq:pEcorr})
by Lorentzians with half width $t$ at half maximum. A dielectric
constant of $\varepsilon\approx10$ was used for both shells. The
dashed line is the fit to the full line with a Gaussian distribution.
The inset shows the true discrete Mn-site energy probability.}

\end{figure}

It is easy to generalise the first NN result $\Delta\epsilon_{\mathrm{Mn}}=4V_{1}$
in order to account for the Coulomb contribution of the the $i\mbox{th}$
shell, $V_{i}=e^{2}/(4\pi\varepsilon_{0}\varepsilon_{i}R_{i})$, where
$\varepsilon_{i}$ is the dielectric constant for the given shell.
In particular, taking into account second and third NNs, we get $\Delta\epsilon_{\textrm{Mn}}=4V_{1}-12V_{3}$,
where $V_{2}$ is absent because the two inequivalent Mn sites have
the same second NN environment. The value $\Delta\epsilon_{\mathrm{Mn}}\approx0.5\,\mathrm{eV}$
found by Pickett and Singh \cite{PS97} is reproduced with $\varepsilon_{1}\approx10$
and $\varepsilon_{3}\approx17$, where we used $R_{3}=a\sqrt{19}/2\approx8.5\,\mbox{\AA}$.
Following \cite{PS97}, we will keep only first and second NN contributions,
with $\varepsilon_{1}=\varepsilon_{2}\approx10$, and $R_{2}=a\sqrt{11}/2\approx6.5\,\mbox{\AA}$.
The resulting random site energies may be written as \begin{equation}
\epsilon_{i}=V_{1}(l_{i1}+l_{i2}\sqrt{3/11}),\label{eq:eiMang}\end{equation}
where $l_{ij}$ is the number of $AE^{2+}$ ions in the $j$th shell
of Mn site~$i$ for a given $RE^{3+}/AE^{2+}$ configuration. Inserting
$\varepsilon_{1}\approx10$ in equation~(\ref{eq:V1mang}) we get
$V_{1}\approx0.43\,\mbox{eV}\approx2.1t$. The probability distribution
for this random, but correlated site energy model may be written as\begin{eqnarray}
p(\epsilon_{i}) & = & \sum_{l_{i2}=0}^{24}\left(\begin{array}{c}
24\\
l_{i2}\end{array}\right)\sum_{l_{i1}=0}^{8}\left(\begin{array}{c}
8\\
l_{i1}\end{array}\right)x^{l_{i1}+l_{i2}}\times\nonumber \\
 &  & (1-x)^{32-l_{i1}-l_{i2}}\delta\biggl[\epsilon_{i}-V_{1}\Bigl(l_{i1}+l_{i2}\frac{\sqrt{3}}{\sqrt{11}}\Bigr)\biggr].\label{eq:pEcorr}\end{eqnarray}
In figure~\ref{fig:dist2NN} we show the resulting coarse grained
distribution as a function of $\epsilon_{i}-\left\langle \epsilon_{i}\right\rangle $
for $x=1/3$, where the average site energy is $\left\langle \epsilon_{i}\right\rangle =xV_{1}(8+24\sqrt{3/11})$;
the inset shows the true discrete Mn-site energy probability, as given
by the weight of the delta functions in (\ref{eq:pEcorr}). The distribution
is approximately Gaussian with a root mean square (RMS) deviation\begin{equation}
\sigma\equiv\sqrt{\langle\epsilon_{i}^{2}\rangle-\langle\epsilon_{i}\rangle^{2}}\simeq4.6t,\label{eq:rms}\end{equation}
as obtained by fitting with a Gaussian distribution (dashed line in
figure~\ref{fig:dist2NN}). A rectangular distribution with the same
RMS deviation has $W=\sigma\sqrt{12}\approx15.9t$; well in the range
required for a MIT at $T_{\mathrm{C}}$, $12t<W<16.5t$ \cite{SXS+97b,ZMJ05}.
The effect on $T_{\mathrm{C}}$ of a discrete random site-energy distribution
similar to equation~(\ref{eq:pEcorr}), but restricted to the NN
shell, has been considered in \cite{SB06}.

The analysis which lead us to figure~\ref{fig:dist2NN} and equation~(\ref{eq:rms})
was previously carried out in \cite{PS97}, but with a very different
conclusion about the strength of disorder in CMR manganites. The main
difference with respect to the present analysis is the value of the
dielectric constant entering equation~(\ref{eq:pEcorr}) through
$V_{1}$%
\footnote{An additional difference comes from the misplaced second shell of
$RE^{3+}/AE^{2+}$ sites in \cite{PS97}. The distances to the first
and to the second shells differ by 48\%, and not by the referred 13\%
in \cite{PS97}. This error, however, does not change considerably
the results: while in the case of 48\% difference the second shell
has 24 sites, in the case of 13\% difference it has only 6 sites.%
}. Even though it is stated in \cite{PS97} that first and second shells
of $RE^{3+}/AE^{2+}$ sites are taken into account with $\varepsilon_{1}=\varepsilon_{2}\approx10$,
we can only reproduce figure~1 of \cite{PS97} (the analogous of
our figure~\ref{fig:dist2NN}) if equation~(\ref{eq:pEcorr}) is
used with $\varepsilon_{1}=\varepsilon_{2}\approx34$ (an unlikely
dielectric constant, as discussed above). As a consequence, the associated
distribution was found to have a RMS of $\sigma\approx1.3t\approx0.26\,\mbox{eV}$%
\footnote{In \cite{PS97} the full width of the distribution at half maximum
was found to be $\delta\epsilon\approx0.6\,\mbox{eV}$. If we assume
the distribution to be Gaussian, the full width at half maximum is
related with the root mean square $\sigma$ as $\delta\epsilon=\sigma2\sqrt{2\ln2}$.%
}. This means that a rectangular distribution with the same RMS deviation
has $W=\sigma\sqrt{12}\approx4.5t$; well below the values required
for a MIT at $T_{\mathrm{C}}$, $12t<W<16.5t$ \cite{SXS+97b,ZMJ05}.

\section{Results and discussion}

From the above considerations we may conclude that a realistic parametrisation
of diagonal disorder is mandatory for a precise estimate of the effect
of substitutional disorder in manganites. We have performed a transfer
matrix calculation \cite{MK81,MK83,KM93} using the model given in
equation~(\ref{eq:HDEdis}), with site energies calculated from a
random distribution of the dopant ions $AE^{2+}$. First and second
shells of $RE^{3+}/AE^{2+}$-sites were taken into account assuming
equal dielectric constant, which results in random on-site energies
given by (\ref{eq:eiMang}), and a site energy distribution given
by (\ref{eq:pEcorr}). In this case the site disorder is parametrised
by $x$, which determines the fraction of $AE^{2+}$ ions in the system
and thus the variables $l_{ij}$ in \ref{eq:eiMang}, and by the parameter
$V_{1}$ given in equation~\ref{eq:V1mang}, or equivalently the
dielectric constant $\varepsilon$. The doping level $x$ also determines
the Fermi energy in the system, which was calculated by integrating
over the disorder averaged density of states obtained for clusters
of $64\times64\times64$ sites using the recursive Green's function
method \cite{Hayd80}.

\begin{figure}
\begin{centering}
\includegraphics[clip,width=0.55\columnwidth]{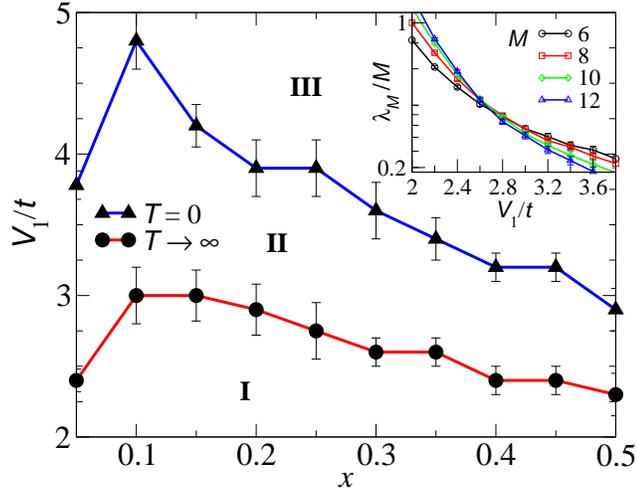}
\par\end{centering}

\caption{\label{cap:Correlated_model-mobility-edge-V-x}Critical value of $V_{1}$
vs $x$ in the paramagnetic (circles) and ferromagnetic phases (triangles).
A system with $(x,V_{1})$ in region \textbf{I (III)} is metallic
(insulator) in the paramagnetic and ferromagnetic phases; in region
\textbf{II} the mobility edge crosses the Fermi level in the para-ferromagnetic
transition. The inset shows the typical behaviour of the normalised
finite-size localisation length $\lambda_{M}/M$ vs $V_{1}/t$ across
the MIT for the particular case of $x=0.30$ in the paramagnetic phase.}

\end{figure}

The transfer matrix technique gives reliable information about the
extended or localised nature of the eigenstates \cite{KM93,SO99}.
For a quasi-one-dimensional system with length $L$ and cross-section
$M\times M$, where $L\gg M$ (in units of the lattice constant),
the method provides the localisation length $\lambda_{M}$ of the
finite system from the smallest Lyapunov exponent of the respective
transfer matrix product. The scaling behaviour of the normalised finite-size
localisation length $\lambda_{M}/M$ then determines the nature of
the eigenstates for a given $x$ and $V_{1}$ at the Fermi energy.
Extended (localised) states show increasing (decreasing) $\lambda_{M}/M$
as $M$ increases. This behaviour can be appreciated in the inset of
figure~\ref{cap:Correlated_model-mobility-edge-V-x} for $x=0.30$
in the paramagnetic phase, where the direction of each spin is chosen
randomly from a uniform distribution on a sphere. We have used system
bars ($L\times M\times M$) with a longitudinal length $L$ such that
the relative error in $\lambda_{M}$ is $\lesssim1\%$ (typically
$L\sim10^{5}$), and $M=6,8,10,12$. At criticality $\lambda_{M}/M$
is independent of $M$, signalling the Anderson transition, and providing
the critical parameter values \cite{XXE+03}.

The main result of this work is presented in figure~\ref{cap:Correlated_model-mobility-edge-V-x}.
For each concentration the critical values of $V_{1}$, at which the
mobility edge and the Fermi level coincide, were calculated in the
paramagnetic (circles) and ferromagnetic phases (triangles). A value
of $V_{1}$ between these two (i.e., in region \textbf{II}) implies
a crossing of the Fermi level and the mobility edge when the system
orders magnetically. A value of $V_{1}\approx3t$ is sufficient to
give rise to an Anderson MI transition for concentrations $x\sim0.2-0.5$.
While it is still higher than the estimate based on \cite{PS97} and
\cite{PS96}, $V_{1}\approx2.1t$, it is sufficiently close to cast
some doubt on a straightforward dismissal of a role of Anderson localisation
in the magnetoresistance of the manganites. Moreover, the value $V_{1}\approx2.1t$
only takes into account the random distribution of potential sources,
namely $RE^{3+}$ and $AE^{2+}$ ions. The presence of $RE/AE$ ionic
size mismatch is expected to enhance considerably the disorder effects
\cite{SST99,AFG+02,KK07}. Note also that, in this model the critical
value of disorder does not vary monotonically with $x$ and shows
a maximum at around $x\sim0.1$. One should bear in mind that, in
this model, changing $x$ also changes the distribution of site energies
{[}equation.~(\ref{eq:pEcorr}){]}, and so $V_{1}$ does not, by
itself, characterise the disorder.

A final remark regarding the relevance of the model used in this work
is in place. The model given by equation~(\ref{eq:HDEdis}) with
correlated on-site disorder as in~(\ref{eq:eiMang}) is certainly
incomplete: it neglects electron-lattice coupling, orbital degrees
of freedom, anti-ferromagnetic exchange between localised $t_{2g}$
spins, and electron-electron interactions between $e_{g}$ electrons.
Consequently, it does not distinguish between the three main manganite
groups (large-, intermediate-, and low-bandwidth) and cannot reproduce
many of the complex phases appearing in each of their phase diagrams
\cite{DHM01}. However, it is general enough and not tremendously
complex to address the question of how important is substitutional
disorder in manganites; this is the motivation for using it.

\section{Conclusions}

We have shown that a careful analysis of the Mn-site energies arising
from random distribution of $AE^{2+}$ and $RE^{3+}$ ions in manganites
produce a probability distribution with a RMS deviation $\sigma\approx4.6t$.
This RMS is a consequence of a parametrisation of screened Coulomb
energies for which an energy shift $V_{1}\approx2.1t$ is assumed
to show up in a Mn-site whenever a NN replacement $RE^{3+}\to AE^{2+}$
occurs. Such a RMS value already places the system in the disorder
window for which a MIT occurs when the ferro-paramagnetic transition
takes place. We have developed a DE model with realistic parametrisation
of on-site disorder which shows to undergo a MIT at the ferro-paramagnetic
transition for $V_{1}\approx3t$. This $V_{1}$ value is slightly
larger than the expected $V_{1}\approx2.1t$, but it is sufficiently
close to unveil the importance of substitutional disorder in manganites
and show that disorder must be considered at least on the same foot
as the coupling to the lattice. These findings give support to theories
where disorder is a key ingredient \cite{NV01}, as the phase-competition
scenario for CMR, where diagonal disorder has been found to play a
crucial role \cite{MMF+00,BMM+01,BMD04,MFN03,MFN05,SAA+06,KM06,SAD07}.
Such a key role played by disorder was also recently observed in a
series of experiments in half-doped manganites \cite{AUT+03,MAA+04},
where A-site ordered and disordered systems have been successfully
prepared and compared.

\ack{}{}

The authors acknowledge financial support from POCI 2010 via project
PTDC/FIS/64404/2006. EVC additionally acknowledges the Juan de la
Cierva Program (MCI, Spain).


\bibliographystyle{iopart-num}

\begin{thebibliography}{10}
\expandafter\ifx\csname url\endcsname\relax
  \def\url#1{{\tt #1}}\fi
\expandafter\ifx\csname urlprefix\endcsname\relax\def\urlprefix{URL }\fi
\providecommand{\eprint}[2][]{\url{#2}}

\bibitem{CVvM99}
Coey J~M~D, {V}iret M and {V}on {M}olnar S 1999 {\em Adv. {P}hys.\/} {\bf 48}
  167 -- 293

\bibitem{LP00}
Loktev V~M and {P}ogorelov Y~G 2000 {\em Low {T}emp. {P}hys.\/} {\bf 26} 171 --
  193

\bibitem{SJ01}
Salamon M~B and {J}aime M 2001 {\em Rev. {M}od. {P}hys.\/} {\bf 73} 583 -- 628

\bibitem{DHM01}
Dagotto E, {H}otta T and {M}oreo A 2001 {\em Phys. {R}ep.\/} {\bf 344} 1 -- 153

\bibitem{Edw02}
Edwards D~M 2002 {\em Adv. {P}hys.\/} {\bf 51} 1259 -- 1318

\bibitem{Dag05}
Dagotto E 2005 {\em Science\/} {\bf 309} 257 -- 262

\bibitem{ZEN51b}
Zener C 1951 {\em Phys. Rev.\/} {\bf 82} 403 -- 405

\bibitem{AH55}
Anderson P~W and Hasegawa H 1955 {\em Phys. Rev.\/} {\bf 100} 675 -- 681

\bibitem{DeGennes60}
de~Gennes P~G 1960 {\em Phys. Rev.\/} {\bf 118} 141 -- 154

\bibitem{KA88}
Kogan E~M and Auslender M~I 1988 {\em phys. stat. sol. (b)\/} {\bf 147} 613 --
  620

\bibitem{AKT88}
Auslender M~I, Kogan E~M and Tretyakov S~V 1988 {\em phys. stat. sol. (b)\/}
  {\bf 148} 289 -- 295

\bibitem{Var96}
Varma C~M 1996 {\em Phys. Rev. B\/} {\bf 54} 7328 -- 7333

\bibitem{KAK99}
Kogan E, Auslender M and Kaveh M 1999 {\em Eur. Phys. J. B\/} {\bf 9} 373--376

\bibitem{LZB+97}
Li Q, Zang J, Bishop A~R and Soukoulis C~M 1997 {\em Phys. Rev. B\/} {\bf 56}
  4541

\bibitem{SXS+97b}
Sheng L, Xing D~Y, Sheng D~N and Ting C~S 1997 {\em Phys. Rev. B\/} {\bf 56}
  R7053 -- R7056

\bibitem{SXS+97a}
Sheng L, Xing D~Y, Sheng D~N and Ting C~S 1997 {\em Phys. Rev. Lett.\/} {\bf
  79} 1710 -- 1713

\bibitem{ZMJ05}
Zar{\'a}nd G, Moca C~P and Jank{\'o} B 2005 {\em Phys. Rev. Lett.\/} {\bf 94}
  247202

\bibitem{PS97}
Pickett W~E and Singh D~J 1997 {\em Phys. Rev. B\/} {\bf 55} R8642 -- R8645

\bibitem{MLS95}
Millis A~J, Littlewood P~B and Shraiman B~I 1995 {\em Phys. Rev. Lett.\/} {\bf
  74} 5144 -- 5147

\bibitem{MSM96}
Millis A~J, Shraiman B~I and Mueller R 1996 {\em Phys. Rev. Lett.\/} {\bf 77}
  175 -- 178

\bibitem{RZB96}
R{\"o}der H, Zang J and Bishop A~R 1996 {\em Phys. Rev. Lett.\/} {\bf 76} 1356
  -- 1359

\bibitem{BSD+06}
Bozin E~S, Schmidt M, DeConinck A~J, Paglia G, Mitchell J~F, Chatterji T,
  Radaelli P~G, Proffen T and Billinge S~J~L 2007 {\em Phys. Rev. Lett.\/} {\bf
  98} 137203

\bibitem{MMF+00}
Moreo A, Mayr M, Feiguin A, Yunoki S and Dagotto E 2000 {\em Phys. Rev.
  Lett.\/} {\bf 84} 5568 -- 5571

\bibitem{BMM+01}
Burgy J, Mayr M, Martin-Mayor V, Moreo A and Dagotto E 2001 {\em Phys. Rev.
  Lett.\/} {\bf 87} 277202

\bibitem{BMD04}
Burgy J, Moreo A and Dagotto E 2004 {\em Phys. Rev. Lett.\/} {\bf 92} 097202

\bibitem{MFN03}
Motome Y, Furukawa N and Nagaosa N 2003 {\em Phys. Rev. Lett.\/} {\bf 91}
  167204

\bibitem{MFN05}
Motome Y, Furukawa N and Nagaosa N 2005 {\em Lecture Notes in Physics\/} vol
  678 ed Donath M and Nolting W (Springer-Verlag) p~71

\bibitem{KM06}
Kumar S and Majumdar P 2006 {\em Phys. Rev. Lett.\/} {\bf 96} 016602

\bibitem{SAD07}
Sen C, Alvarez G and Dagotto E 2007 {\em Phys. Rev. Lett.\/} {\bf 98} 127202

\bibitem{SAA+06}
Sen C, Alvarez G, Aliaga H and Dagotto E 2006 {\em Phys. Rev. B\/} {\bf 73}
  224441

\bibitem{PS96}
Pickett W~E and Singh D~J 1996 {\em Phys. Rev. B\/} {\bf 53} 1146 -- 1160

\bibitem{NV01}
Narimanov E~E and Varma C~M 2001 {\em Phys. Rev. B\/} {\bf 65} 024429

\bibitem{MK81}
Mackinnon A and {K}ramer B 1981 {\em Phys. {R}ev. {L}ett.\/} {\bf 47} 1546 --
  1549

\bibitem{MK83}
Mackinnon A and {K}ramer B 1983 {\em Z. {P}hys. {B} - {C}ondens. {M}atter\/}
  {\bf 53} 1 -- 13

\bibitem{KM93}
Kramer B and MacKinnon A 1993 {\em Rep. Prog. Phys.\/} {\bf 56} 1469--1564

\bibitem{CVR+95}
Coey J~M~D, Viret M, Ranno L and Ounadjela K 1995 {\em Phys. Rev. Lett.\/} {\bf
  75} 3910 -- 3913

\bibitem{BKB+99}
Boris A~V, Kovaleva N~N, Bazhenov A~V, van Bentum P~J~M, Rasing T, Cheong S~W,
  Samoilov A~V and Yeh N~C 1999 {\em Phys. Rev. B\/} {\bf 59} R697 -- R700

\bibitem{SB06}
Salafranca J and Brey L 2006 {\em Phys. Rev. B\/} {\bf 73} 214404

\bibitem{Hayd80}
Haydock R 1980 {\em Solid State Physics\/} vol~35 ed Ehrenreich H, Seitz F and
  Turnbull D (New York: Academic Press) p 215

\bibitem{SO99}
Slevin K and {O}htsuki T 1999 {\em Phys. {R}ev. {L}ett.\/} {\bf 82} 382 -- 385

\bibitem{XXE+03}
Xiong S~J, Xing D~Y, Evangelou S~N and Sheng D~N 2003 {\em Phys. Lett. A\/}
  {\bf 311} 426 -- 431

\bibitem{SST99}
Sheng L, Sheng D~N and Ting C~S 1999 {\em Phys. Rev. B\/} {\bf 59} 13550 --
  13553

\bibitem{AFG+02}
Alonso J~L, {F}ernandez L~A, {G}uinea F, {L}aliena V and {M}artin {M}ayor V
  2002 {\em Phys. {R}ev. {B}\/} {\bf 66} 104430

\bibitem{KK07}
Kumar S and Kampf A~P 2008 {\em Phys. Rev. Lett.\/} {\bf 100} 076406

\bibitem{AUT+03}
Akahoshi D, {U}chida M, {T}omioka Y, {A}rima T, {M}atsui Y and {T}okura Y 2003
  {\em Phys. {R}ev. {L}ett.\/} {\bf 90} 177203

\bibitem{MAA+04}
Mathieu R, {A}kahoshi D, {A}samitsu A, {T}omioka Y and {T}okura Y 2004 {\em
  Phys. {R}ev. {L}ett.\/} {\bf 93} 227202

\end{thebibliography}

\providecommand{\newblock}{}

\end{document}